\documentclass[aps,showpacs,nofootinbib,twocolumn]{revtex4}
\usepackage{amsmath}
\usepackage{mathtools}
\usepackage{amssymb}
\usepackage{mathrsfs}
\usepackage{epsfig,amssymb,amsmath}
\usepackage{subfigure}
\usepackage{CJK}
\usepackage[english]{babel}

\newtheorem{theorem}{Theorem}
\newtheorem{proposition}[theorem]{Proposition}

\begin{document}
\title{Classification of the Entangled States of $2\times L\times M\times N\times H$}
\author{Kang-Kang Jia$^1$}
\author{Jun-Li Li$^1$}
\author{Cong-Feng Qiao$^{1,2}$\footnote{Corresponding author, qiaocf@ucas.ac.cn}}
\affiliation{$^1$School of Physics, University of Chinese Academy of Sciences, YuQuan Road 19A, Beijing 100049, China\\
$^2$CAS Center for Excellence in Particle Physics, Beijing 100049, China}

\begin{abstract}
In this work we propose a practical entanglement classification scheme for
pure states of $2\times L\times M\times N\times H$, under the stochastic
local operation and classical communication (SLOCC), which generalizes the
method explored in the entanglement classification of $2\times L\times
M\times N$ to the five-partite system. The entangled states of $2\times
L\times M\times N\times H$ system are first classified into different
coarse-grained standard forms using matrix decompositions, and then
fine-grained identification of two inequivalent entangled states with the
same standard form are completed by using the matrix realignment
technique. As an practical example, entanglement classes of the five-qubit
system of $2\times 2\times 2\times 2\times 2$ are presented.
\end{abstract}
\pacs{03.65.Ud, 03.67.Bg, 03.67.Mn}
\maketitle

\section{Introduction}

Entanglement is an essential feature of quantum theory, and now has been
considered to be the key physical resource of quantum information sciences.
Many nonclassical applications can only be implemented when entangled states
are explored, e.g., quantum teleportation \cite{quantum-tel}, dense coding
\cite{dense-coding92,dense-coding96}, and some of the quantum cryptography
protocols \cite{crypto-bell}. However, many superficially different quantum
states may have actually the same function when being applied to carry out
the quantum information tasks. It is known that, if two entangled states are
interconnected by invertible local operators, i.e., equivalent under
stochastic local operation and classical communication (SLOCC), then they
would be both applicable for the same quantum information tasks. While there
are only two SLOCC inequivalent tripartite entanglement classes in
three-qubit system \cite{three-qubit}, the inequivalent classes turn to
infinite when the system consists more than three partite.

The entanglement classification under SLOCC is generally a difficult problem
as the particles and dimensions of each partite grows, though it would be
much easier when the entangled states has particular symmetries
\cite{symmetric N-qubit states}. At present, nine inequivalent families of
quantum systems for four-qubit states under SLOCC have been identified due
to the symmetric property SU(2)$\otimes$SU(2)$\simeq$SO(4)
\cite{four-qubit-nine}. Finer grained classifications could also be achieved
with well constructed entangled measures \cite{anti-linear1,anti-linear2}.
Using a technique of coefficient matrix \cite{coefficient-matrix}, 28
genuinely entangled families were found for the four-qubit system
\cite{four-qubit-28}. The rank of the coefficient matrix is useful in
partitioning the entangled states into discrete entanglement families
\cite{general-coefficient-matrix}, however as the dimensions and number of
particles both grow, it provides a rather coarse grained classification
\cite{222d}. New method for the entanglement classification of $2\times
L\times M\times N$ system has been proposed \cite{2lmn} which takes full
advantage of the classifications of $2\times M\times N$ system \cite{2nn,
2mn,2mn-parameters,LNN}. The method not only provides a even finer
classification for the system, but also is capable of determining the
equivalency of two quantum states falling into the same entanglement family.

In this work, we generalize the method in \cite{2lmn} to the case of
five-partite system of $2\times L\times M\times N\times H$. The five-partite
system with one qubit is first partitioned into tri-partite in form of
$2\times (L\times M)\times (N\times H)$, and the standard forms of
inequivalent entanglement classes of $2\times (L M)\times (N H)$ behave as
the entanglement families of $2\times L\times M\times N\times H$. Then the
matrix realignment is utilized to determine the equivalence of two entangled
states and the connecting matrices between them within the same family. The
content goes as follows, in Sec.II, the classification procedures of
$2\times L\times M\times N\times H$ are presented. In Sec.III, the
classification of the five-qubit system is given as a concrete example,
where detailed comparisons with the results in literature are also
presented. Summary are conclued in Sec.IV.

\section{The entanglement classification of pure system of
$2\times L\times M\times N\times H$}

\subsection{The representation of five-partite states}

Every quantum state $|\psi\rangle$ of five-partite system $2\times L\times
M\times N\times H$ may be formulated as the following
\begin{eqnarray}
|\psi\rangle = \sum_{i,m,n,l,h=1}^{2,M,N,L,H} \gamma_{imnlh} |i,m,n,l,h\rangle \; ,
\end{eqnarray}
where $\gamma_{ilmnh} \in \mathbb{C}$ are coefficients of the state in
representative bases. Therefore, the quantum state $|\psi\rangle$ may also
be represented as a high dimensional complex tensor $\psi$ whose matrix
elements are $\gamma_{ilmnh}$. In this form, the SLOCC equivalence of two
quantum states $\psi'$ and $\psi$ may be formulated as \cite{three-qubit}
\begin{eqnarray}
\psi' = A^{(1)} \otimes A^{(2)} \otimes A^{(3)} \otimes
 A^{(4)} \otimes A^{(5)} \psi \; , \label{Psi-A-SLOCC}
\end{eqnarray}
here $A^{(1)} \in \mathbb{C}^{2\times 2}$,  $A^{(2)} \in \mathbb{C}^{L\times
L}$, $A^{(3)} \in \mathbb{C}^{M\times M}$, $A^{(4)} \in \mathbb{C}^{N\times
N}$, $A^{(5)} \in \mathbb{C}^{H\times H}$  are invertible matrices of
$2\times 2$, $L\times L$, $M\times M$, $N\times N$, $H\times H$ separately,
which act on the corresponding particles.

For the sake of clarity, the quantum state $\psi$ may also be
formulated as $\psi \doteq \begin{pmatrix}\Gamma_1 \\
\Gamma_2 \end{pmatrix}$, and
\begin{eqnarray}
\begin{pmatrix}\Gamma_{1} \\ \Gamma_{2}\end{pmatrix}=
\begin{pmatrix} \begin{pmatrix}
\gamma_{11111} & \gamma_{11112} & \cdots & \gamma_{111NH} \\
\gamma_{11211} & \gamma_{11212} & \cdots & \gamma_{112NH} \\
\vdots & \vdots & \ddots & \vdots \\
\gamma_{1LM11} & \gamma_{1LM12} & \cdots & \gamma_{1LMNH}
\end{pmatrix} \\ \\
\begin{pmatrix}
\gamma_{21111} & \gamma_{21112} & \cdots & \gamma_{211NH} \\
\gamma_{21211} & \gamma_{21212} & \cdots & \gamma_{212NH} \\
\vdots & \vdots & \ddots & \vdots \\
\gamma_{2LM11} & \gamma_{2LM12} & \cdots & \gamma_{2LMNH}
\end{pmatrix} \end{pmatrix} \; ,  \label{Four-Matrix-pair}
\end{eqnarray}
which is obtained by grouping the particles as $2 \times (L\times M)\times
(N\times H)$. Here $\Gamma_{i}\in \mathbb{C}^{LM\times NH}$, i.e. complex
matrices of $LM$ columns and $NH$ rows (we may assume $LM \leq NH$ without
loss of generalities).

\subsection{The entanglement families of $2\times L\times M\times N\times H$ system}

It is easy to observe that the quantum state of tripartite system of
$2\times LM\times NH$ could also be represented in same form as
Eq.(\ref{Four-Matrix-pair}). Following the method introduced in \cite{2lmn},
the SLOCC equivalence of two states $\psi'$ and $\psi$ in Eq.
(\ref{Psi-A-SLOCC}) transforms into the following form
\begin{eqnarray}
\psi' = T\otimes P \otimes Q^{\mathrm{T}}\psi \; ,
\end{eqnarray}
and in the matrix pair representations, we have
\begin{eqnarray}
\begin{pmatrix}
\Gamma_1' \\ \Gamma_2'
\end{pmatrix} = A^{(1)}
\begin{pmatrix}
P \Gamma_1 Q \\ P \Gamma_2 Q
\end{pmatrix} \; ,  \label{Four-Gamma}
\end{eqnarray}
here $P = A^{(2)}\otimes A^{(3)}$, $Q^{\mathrm{T}} = A^{(4)}\otimes
A^{(5)}$, $\mathrm{T}$ stands for matrix transposition, $A^{(1)}$ acts on
the two matrices $\Gamma_{1,2}$, and $P$ and $Q$ act on the rows and columns
of the $\Gamma_{1,2}$ matrices, accordingly. The SLOCC equivalence of two
$2\times L\times M\times N\times H$ quantum states in Eq.(\ref{Four-Gamma})
has a similar form to the tripartite $2\times LM\times NH$ pure state
\cite{2mn}. The differences lies in that $P$ and $Q$ are not only invertible
operators but also direct products of two invertible matrices, $A^{(2)}$ and
$A^{(3)}$, $A^{(4)}$ and $A^{(5)}$.

Similar as that of \cite{2lmn}, we have the following proposition:
\begin{proposition}
If two quantum states of\, $2\times L\times M\times N\times H$ are SLOCC
equivalent then their corresponding matrix-pairs have the same standard
forms as that of $2\times LM\times NH$ under the invertible operators $T\in
\mathbb{C}^{2\times 2}$, $P\in \mathbb{C}^{LM\times LM}$, $Q\in
\mathbb{C}^{NH\times NH}$. \label{Proposition-standard-forms}
\end{proposition}

This proposition serves as a necessary condition for the SLOCC equivalence
of the entangled states of the $2\times L\times M\times N\times H$ system.

The transforming matrices $T_0$,  $P_0$, $Q_0$ for the standard form can be
obtained. Generally the transformation matrices for the standard form are
not unique. For example, if $T_0$, $P_0$, $Q_0$ are the matrices that
transform $\psi$ into its standard form, then the following matrices will do
likewise
\begin{eqnarray}
T_0 \otimes SP_0 \otimes (Q_0 S^{-1})^{\mathrm{T}} \psi =
\begin{pmatrix} E \\ J \end{pmatrix} \; , \label{Matrix-pair-invariant}
\end{eqnarray}
where $SJS^{-1} = J$, i.e. $[S,J]=0$. The nonuniqueness comes from the
symmetries of standard forms.

\subsection{The entanglement classification of a $2\times L\times M\times N\times H$ system}

As the main result of the paper, we present the following theorem
\begin{theorem}
Two $2\times L\times M\times N\times H$ quantum states $\psi$ and $\psi'$
are SLOCC equivalent if and only if their corresponding matrix-pair
representations have the same standard forms of $2\times LM\times NH$ and
the transformation matrices $P$ and $Q$ in Eq.(\ref{Four-Gamma}) have the
form of direct products of two invertible matrices, i.e., $P =
A^{(2)}\otimes A^{(3)}$ and $Q^{\mathrm{T}} = A^{(4)}\otimes A^{(5)}$.
\label{theorem-2lmn}
\end{theorem}
\noindent {\bf Proof:} If two $2\times L\times M\times N\times H$ quantum
states $\psi$ and $\psi'$ are SLOCC equivalent, we have
\begin{eqnarray}
\psi' = A^{(1)} \otimes A^{(2)} \otimes A^{(3)} \otimes A^{(4)} \otimes A^{(5)} \; \psi
\; , \label{psi-psip-1}
\end{eqnarray}
here $A^{(i)}$ is invertible matrix, $i\in \{1,2,3,4,5\}$. According to
Proposition \ref{Proposition-standard-forms}, we have
\begin{eqnarray}
\psi' = T\otimes P \otimes Q^{\mathrm{T}} \; \psi \; , \label{psi-psip-2}
\end{eqnarray}
which means that $\psi'$ and $\psi$ have the same standard form of $2\times
LM\times NH$. Combining Eq.(\ref{psi-psip-1}) and Eq.(\ref{psi-psip-2})
yields
\begin{eqnarray}
&&T^{-1}A^{(1)} \otimes (P^{-1}(A^{(2)} \otimes A^{(3)}))\otimes  \notag \\
&&((Q^{\mathrm{T}})^{-1} A^{(4)} \otimes A^{(5)}) \psi= \psi \; .
\end{eqnarray}
As the unit matrices $E\otimes E\otimes E$ must be one of the operators
which stabilizes the quantum state $\psi$ in the matrix-pair form, $P$ and
$Q^{\mathrm{T}}$ have the solution of $P = A^{(2)}\otimes A^{(3)}$ and
$Q^{\mathrm{T}} = A^{(4)}\otimes A^{(5)}$.

If the two quantum states have the same standard form, then we will have
Eq.(\ref{psi-psip-2}). And if further $P$ and $Q$ have the decomposion of $P
= P_1\otimes P_2$ and $Q = Q_1\otimes Q_2$ where $P_1\in \mathbb{C}^{L\times
L}$, $P_2 \in \mathbb{C}^{M\times M}$ and $Q_1\in \mathbb{C}^{N\times N}$,
$Q_2 \in \mathbb{C}^{H\times H}$. As matrices $P$ and $Q$ are invertible if
and only if both $P_1$ and $P_2$, $Q_1$ and $Q_2$ are invertible, thus
\begin{eqnarray}
\psi' = T\otimes (P_1\otimes P_2)\otimes (Q_1\otimes Q_2)^{\mathrm{T}}\; \psi \; .
\end{eqnarray}
Therefore $\psi'$ and $\psi$ are SLOCC equivalent entangled states of a
$2\times L\times M\times N\times H$ system. Q.E.D.

Thus the classification procedure may stated as follows. First, we
constructed the standard forms of the $2\times LM\times NH$ system, which
behave as the entanglement families of $2\times L\times M\times N\times H$
and the transforming matrices $T_0$, $P_0$, $Q_0$ are also obtained. If two
quantum states transform into different families, they are SLOCC
inequivalent. Otherwise, the connecting matrices of $T$, $P$, $Q$ may be
obtained. And we can determine whether such matrices have the direct
products form or not using the matrix realignment technique \cite{2lmn}.
Finally the theorem \ref{theorem-2lmn} provides the complete entanglement
classification for the two entangled states. In the following, we give
detailed examples for $2\times 2\times 2\times 2\times 2$ quantum system as
the application of our method.

\section{Entanglement classification of $2\times 2\times 2\times 2\times 2$ system}

There are totally 32 inequivalent families for the genuine $2\times 2\times 2\times 2\times 2$ entangled classes according to our method, the genuine entangled families of $2\times 2\times 2\times 2\times 2$ quantum states are listed as follows. The $\mathcal{N}_f(22222) = 32$ families includes: \\
two families from $2\times 2\times 2$ system (GHZ and W),
\begin{eqnarray}
&&|\psi\rangle = |1(11)(11)\rangle + |2(22)(22)\rangle \; , \nonumber \\
&&|\psi\rangle = |1(11)(11)\rangle + |1(22)(22)\rangle + |2(11)(22)\rangle \; , \nonumber
\end{eqnarray}
two families from $2\times 2\times 3$ system,
\begin{eqnarray}
|\psi\rangle & = & |1(11)(11)\rangle + |1(12)(12)\rangle + |2(12)(21)\rangle \; , \nonumber \\
|\psi\rangle & = & |1(11)(11)\rangle + |1(12)(12)\rangle + |2(11) (12)\rangle \;  \nonumber \\
&&+ |2(12)(21)\rangle \; , \nonumber
\end{eqnarray}
one family from $2\times 2\times 4$ system,
\begin{eqnarray}
|\psi\rangle & = & |1(11)(11)\rangle + |1(12)(12)\rangle + |2(11)(21)\rangle \;  \nonumber \\
&&+ |2(12)(22)\rangle \; ,\nonumber
\end{eqnarray}
six families from $2\times 3\times 3$ system,
\begin{eqnarray}
|\psi\rangle & = & |1(11)(11)\rangle + |1(12)(12)\rangle + |2(21)(21)\rangle \; , \nonumber \\
|\psi\rangle & = & |1(11)(11)\rangle + |1(12)(12)\rangle + |1(21)(21)\rangle \;   \nonumber \\
&&+ |2(11)(12)\rangle \; , \nonumber \\
|\psi\rangle & = & |1(11)(11)\rangle + |1(12)(12)\rangle + |2(12)(12)\rangle \;   \nonumber \\
&&+ |2(21)(21)\rangle \; , \nonumber \\
|\psi\rangle & = & |1(11)(11)\rangle + |1(12)(12)\rangle + |2(11)(12)\rangle \;   \nonumber \\
&&+ |2(21)(21)\rangle \; , \nonumber \\
|\psi\rangle & = & |1(11)(11)\rangle + |1(12)(12)\rangle + |2(12)(21)\rangle \;   \nonumber \\
&&+ |2(21)(11)\rangle \; , \nonumber \\
|\psi\rangle & = & |1(11)(11)\rangle + |1(12)(12)\rangle + |1(21)(21)\rangle \;   \nonumber \\
&&+ |2(11)(12)\rangle + |2(12)(21)\rangle \; , \nonumber
\end{eqnarray}
five families come from $2\times 3\times 4$ system,
\begin{eqnarray}
|\psi\rangle & = & |1(11)(11)\rangle + |1(12)(12)\rangle + |1(21)(21)\rangle \;   \nonumber \\
&&+ |2(21)(22)\rangle   \; , \nonumber \\
|\psi\rangle & = & |1(11)(11)\rangle + |1(12)(12)\rangle + |1(21)(21)\rangle \;   \nonumber \\
&&+ |2(11)(12)\rangle + |2(21)(22)\rangle \; , \nonumber \\
|\psi\rangle & = & |1(11)(11)\rangle + |1(12)(12)\rangle + |1(21)(21)\rangle \;   \nonumber \\
&&+ |2(11)(11)\rangle + |2(21)(22)\rangle \; , \nonumber \\
|\psi\rangle & = & |1(11)(11)\rangle + |1(12)(12)\rangle + |1(21)(21)\rangle \;   \nonumber \\
&&+ |2(12)(21)\rangle + |2(21)(22)\rangle \; , \nonumber \\
|\psi\rangle & = & |1(11)(11)\rangle + |1(12)(12)\rangle + |1(21)(21)\rangle \;   \nonumber \\
&&+ |2(11)(12)\rangle + |2(12)(21)\rangle + |2(21)(22)\rangle \; . \nonumber
\end{eqnarray}

The other 16 families come from the standard forms of a $2\times 4\times 4$
system. Among the 16 standard forms of $2\times 4\times 4$, there also exist
the continuous entanglement families. That is, different entanglement
families arise from the different values of the characterization parameters.
We have proved that the standard forms with the continuous parameters
belonging to the same entanglement class of $2\times 4\times 4$ system,
correspond to different entanglement families of $2\times 2\times 2\times
2\times 2$ system.

In addition, a necessary condition for the genuine entanglement of a
$2\times L\times M\times N\times H$ system is that all dimensions of the five
particles shall be involved in the entanglement, requiring that $LM \leq
2NH$ with assuming the larger value of the dimensions to be $LM$. The scheme
works better for higher dimensions, especially in the case of $LM = NH$.

\section{Summaries}

In conclusion, we have proposed a practical classification scheme for the
entangled states of $2\times L\times M\times N\times H$ pure system under
SLOCC. By using the standard forms of $2\times LM\times NH$, the entangled
families of $2\times L\times M\times N\times H$ are obtained. And the
invertible local operators that connecting two quantum states in the same
family may also be constructed by using the matrix realignment technique.
This provides a necessary and sufficient condition on the SLOCC equivalence
of the two quantum states. As an application, detailed examples of the
entanglement classification under SLOCC for five-qubit system is presented,
which has not been discussed systematically in the literature to the best of
our knowledge.

\vspace{.7cm}
{\bf Acknowledgments}
We are grateful to Junli Li for discussion. This work was supported in part by National Key Basic Research Program of China under the grant 2015CB856700, and by the National Natural Science Foundation of China(NSFC) under the grants 11175249, 11121092, and 11375200.


\end{document}